\documentclass[prl,superscriptaddress,twocolumn,reprint]{revtex4-1}

\usepackage{amssymb}
\usepackage{natbib}
\usepackage{graphicx}
\usepackage{amsmath}
\usepackage[bookmarks = false]{hyperref}

\begin{document}

\title{Efficient and long-lived quantum memory with cold atoms inside a ring
cavity}
\author{Xiao-Hui Bao}
\affiliation{Physikalisches Institut der Universitaet Heidelberg, Philosophenweg 12,
Heidelberg 69120, Germany}
\affiliation{Hefei National Laboratory for Physical Sciences at Microscale and Department
of Modern Physics, University of Science and Technology of China, Hefei,
Anhui 230026, China}
\author{Andreas Reingruber}
\affiliation{Physikalisches Institut der Universitaet Heidelberg, Philosophenweg 12,
Heidelberg 69120, Germany}
\author{Peter Dietrich}
\affiliation{Physikalisches Institut der Universitaet Heidelberg, Philosophenweg 12,
Heidelberg 69120, Germany}
\author{Jun Rui}
\affiliation{Hefei National Laboratory for Physical Sciences at Microscale and Department
of Modern Physics, University of Science and Technology of China, Hefei,
Anhui 230026, China}
\author{Alexander D\"{u}ck}
\affiliation{Physikalisches Institut der Universitaet Heidelberg, Philosophenweg 12,
Heidelberg 69120, Germany}
\author{Thorsten Strassel}
\affiliation{Physikalisches Institut der Universitaet Heidelberg, Philosophenweg 12,
Heidelberg 69120, Germany}
\author{Li Li}
\affiliation{Hefei National Laboratory for Physical Sciences at Microscale and Department
of Modern Physics, University of Science and Technology of China, Hefei,
Anhui 230026, China}
\author{Nai-Le Liu}
\affiliation{Hefei National Laboratory for Physical Sciences at Microscale and Department
of Modern Physics, University of Science and Technology of China, Hefei,
Anhui 230026, China}
\author{Bo Zhao}
\affiliation{Institute for Theoretical Physics, University of Innsbruck, A-6020
Innsbruck, Austria}
\affiliation{Hefei National Laboratory for Physical Sciences at Microscale and Department
of Modern Physics, University of Science and Technology of China, Hefei,
Anhui 230026, China}
\author{Jian-Wei Pan}
\affiliation{Physikalisches Institut der Universitaet Heidelberg, Philosophenweg 12,
Heidelberg 69120, Germany}
\affiliation{Hefei National Laboratory for Physical Sciences at Microscale and Department
of Modern Physics, University of Science and Technology of China, Hefei,
Anhui 230026, China}
\maketitle

%\pacs{03.67.Hk, 42.50.Ex, 42.50.Nn, 42.50.Dv}

\textbf{Quantum memories are regarded as one of the fundamental building blocks of linear-optical quantum computation \cite{KLM} and long-distance quantum communication \cite{Duan2001}. A long standing goal to realize scalable quantum information processing is to build a long-lived and efficient quantum memory. There have been significant efforts distributed towards this goal. However, either efficient but short-lived \cite{Simon2007, Hedges2010} or long-lived but inefficient quantum memories \cite{Zhao2009, ZhaoR2009, Radnaev2010} have been demonstrated so far. Here we report a high-performance quantum memory in which long lifetime and high retrieval efficiency meet for the first time. By placing a ring cavity around an atomic ensemble, employing a pair of clock states, creating a long-wavelength spin wave, and arranging the setup in the gravitational direction, we realize a quantum memory with an intrinsic spin wave to photon conversion efficiency of 73(2)\% together with a storage lifetime of 3.2(1) ms. This realization provides an essential tool towards scalable linear-optical quantum information processing.}

\vspace{0.5cm}

A high-performance quantum memory is of crucial importance for large-scale
linear-optical quantum computation\cite{KLM}, distributed quantum computing,
and long-distance quantum communication\cite{Duan2001}. The lifetime and the
retrieval efficiency of a quantum memory are two important
quantities that determine the scalability of realistic quantum information
protocols. For a certain quantum information task, e.g. creating a
large-scale cluster state \cite{Browne2005} or distributing entanglement through the quantum repeater protocol \cite{Briegel1998,Bodiya06,Bo07,Sangouard08}, the time overhead $T_{r}$ is inversely proportional to a power law of the retrieval efficiency $R$, $T_{r}\propto R^{-n}$, where $n$ is determined by the scale of the quantum
computation or the communication distance. In order to implement one of those tasks, the lifetime of the quantum memory must be larger than this time overhead. To satisfy this condition, one has to improve the lifetime of the quantum memory and reduce the time overhead by improving the retrieval efficiency. Besides, different protocols also set thresholds on the retrieval efficiency and lifetime. For example, in loss-tolerant linear-optical quantum computation the minimum retrieval efficiency required is 50\% \cite{Rudolph2008} and in long-distance quantum communication distributing entanglement over 1000 km requires a communication time of at least 3.3 ms.

Quantum memories for light have been demonstrated with atomic ensembles \cite
{Kuzmich2005nature,Lukin2005,Polzik2006}, solid state systems \cite
{Tittel2011,Gisin2011}, and single atoms \cite{Rempe2011}. With these quantum memories, the principle of some quantum information protocols have been demonstrated, e.g., functional quantum repeater nodes were realized with atomic ensembles \cite{Chou07,Yuan2008}. However, due to the low retrieval efficiency and short lifetime, the implementation of further steps is extremely difficult. Therefore, in recent years, many efforts have been devoted towards improving the retrieval efficiency and the lifetime of the quantum memories and significant progress has been achieved. However, an efficient and long-lived quantum memory remains still challenging. For example, the lifetime of the atomic quantum memory has been experimentally improved to the millisecond regime either by increasing the spin wave wavelength \cite{Zhao2009} or confining the atoms in an optical lattice \cite{ZhaoR2009,Radnaev2010}, while the intrinsic retrieval efficiency is less than 25\%. Efficient quantum memories have been demonstrated in atomic \cite{Simon2007} and solid-state media \cite{Hedges2010}, reaching retrieval efficiencies up to 84\% and 69\% respectively. However, the observed lifetimes were severely limited (240 ns and 3 $\mu $s respectively). The
incapability of accomplishing long lifetime and high retrieval efficiency in
a single quantum memory strongly limits more advanced operations, such as
creating or purifying entanglement \cite{Pan2001} between two or more
quantum repeater nodes or entangling six atomic ensembles to generate GHZ
states which are the building blocks of linear-optical quantum computing \cite{Barrett2010}, and thus is one of the current bottlenecks hindering the implementation of scalable quantum information processing.

\begin{figure*}[hbtp]
\centering
\includegraphics[width=0.8\textwidth]{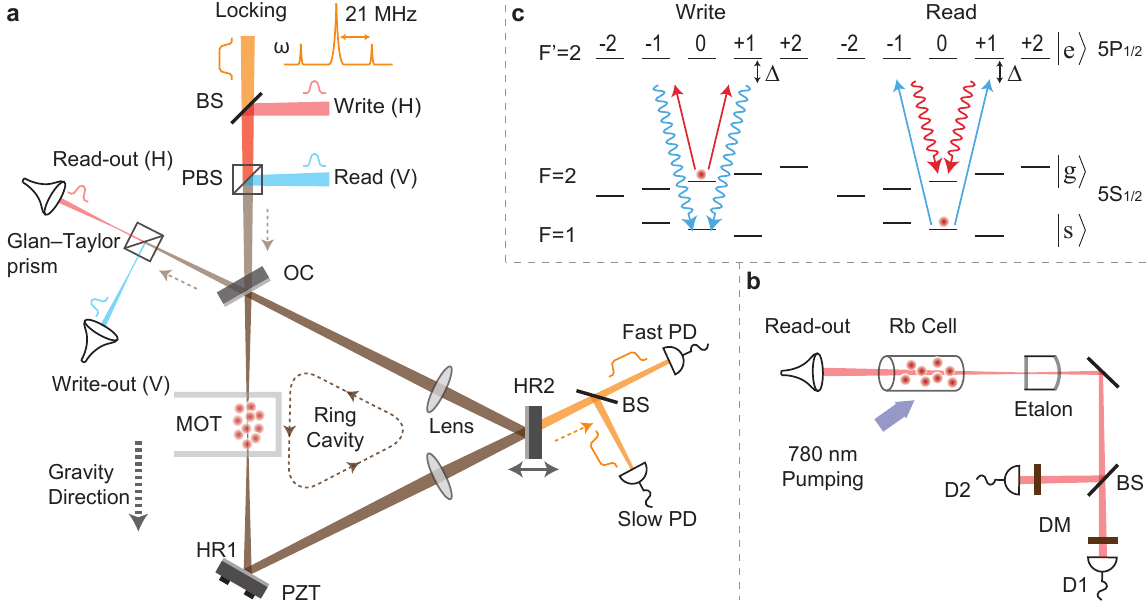}
\caption{\textbf{Experimental setup and level scheme.} \textbf{a}, A ring
cavity is built outside of a vacuum glass cell with antireflection coating.
It mainly consists of two highly reflecting mirrors (HR1 and HR2) and an
output coupler (OC). HR1 is mounted on a piezoelectric transducer (PZT) for
the active stabilization of the cavity length. HR2 is mounted on a manual
translation stage to change the cavity free spectral range.
The write is horizontally (H) polarized and the read is vertically (V) polarized
relative to the ring-cavity plane. These two control beams are combined with a polarizing beamsplitter (PBS) and coupled into the cavity through OC. Write-out and read-out photons are collected in the other port of OC. A frequency modulated locking beam, which is blue detuned by the cavity's free spectral range, is combined with the write through a nonpolarizing beamsplitter (BS). Leakage of the locking beams through HR2 is measured with a fast and a slow photodiode(PD) to create the error signal
for the cavity length stabilization. \textbf{b}, Filtering of read-out photons from
the read beams is realized by making use of a Glan-Taylor prism (in \textbf{a}), a
pumped Rb vapor cell, and an etalon. Dichroic mirrors (DM) are used to remove
the leakage of the vapor cell pumping beams. Finally the read-out photons are
detected with single-photon detectors (D1 and D2). The setup for filtering
write-out photons from write is the same as the read-out channel
(not shown). \textbf{c}, $\Lambda$-type level scheme used. Both the write and read
beams are red detuned by $\Delta=40$ MHz.}
\label{setup}
\end{figure*}

Here, we report an atomic-ensemble quantum memory which has a long
lifetime and a high retrieval efficiency simultaneously. In order to achieve
this challenging goal, we employ a ring cavity to enhance the write and read
processes through the Purcell effect, use a pair of clock states to suppress magnetic
field induced decoherence, and design the directions of the classical beams
and single-photons to be collinear to obtain a spin wave with the maximal
wavelength. We also design the setup and detection beams to be in the
direction of gravity to maintain a good overlap between the control beams
and the atomic ensemble during the free fall. By making these efforts, we
have managed to achieve a quantum memory with an initial intrinsic retrieval efficiency of 73(2)\% and a $1/e$ lifetime of 3.2(1) ms.

The experimental setup is shown in Fig. \ref{setup}. An atomic ensemble composed of $%
\sim 10^{8}$ $\mathrm{^{87}Rb}$ atoms, prepared by a magneto-optical trap
with subsequent molasses cooling, serves as quantum memory. Initially, the
atoms are in the ground state $|g\rangle $. During the write process a
single-quanta spin wave is imprinted on the atomic ensemble conditioned on
the detection of a spontaneously emitted write-out photon (in the same spatial
mode as the write beam). By applying the read beam after a controllable
delay, the spin wave is converted back into a read-out photon. To improve
the retrieval efficiency, a ring cavity with a finesse of 48(1) is placed
around the vacuum chamber to increase the effective optical depth. The
cavity mode is aligned to overlap with the center of the atomic ensemble.
The use of a ring cavity allows us to distinguish the back scattered from
forward scattered write-out photons, so that we can select only the forward
scattered photons and obtain a long-wavelength spin wave which is robust
against dephasing caused by the atomic random motion.

In our experiment, we design the system in such a way that the frequency and
polarization of the write(read) and read(write)-out photons are the same and
that the cavity supports and enhances all the four light fields. We stabilize the cavity intermittently using the Pound-Drever-Hall locking scheme during the MOT loading phase. The phase shift of the write pulse and the read-out photon caused by the atomic
ensemble is compensated by slightly shifting the locking point. Leakage of
the classical write and read pulses into the single-photon channels are
attenuated by a factor of $\sim 10^{10}$ with three stages of filtering
elements shown in Fig. \ref{setup}a and b. Since the write and read are
configured to be co-propagating, a very high extinction ratio ($>10^{11}$)
is required for the pulse switching of the write, read and the locking beam.
This is realized by making use of a double-pass acousto-optic modulator (AOM) and a single-path AOM together for each of these three beams.

\begin{figure}[hbtp]
\centering
\includegraphics[width=0.8\columnwidth]{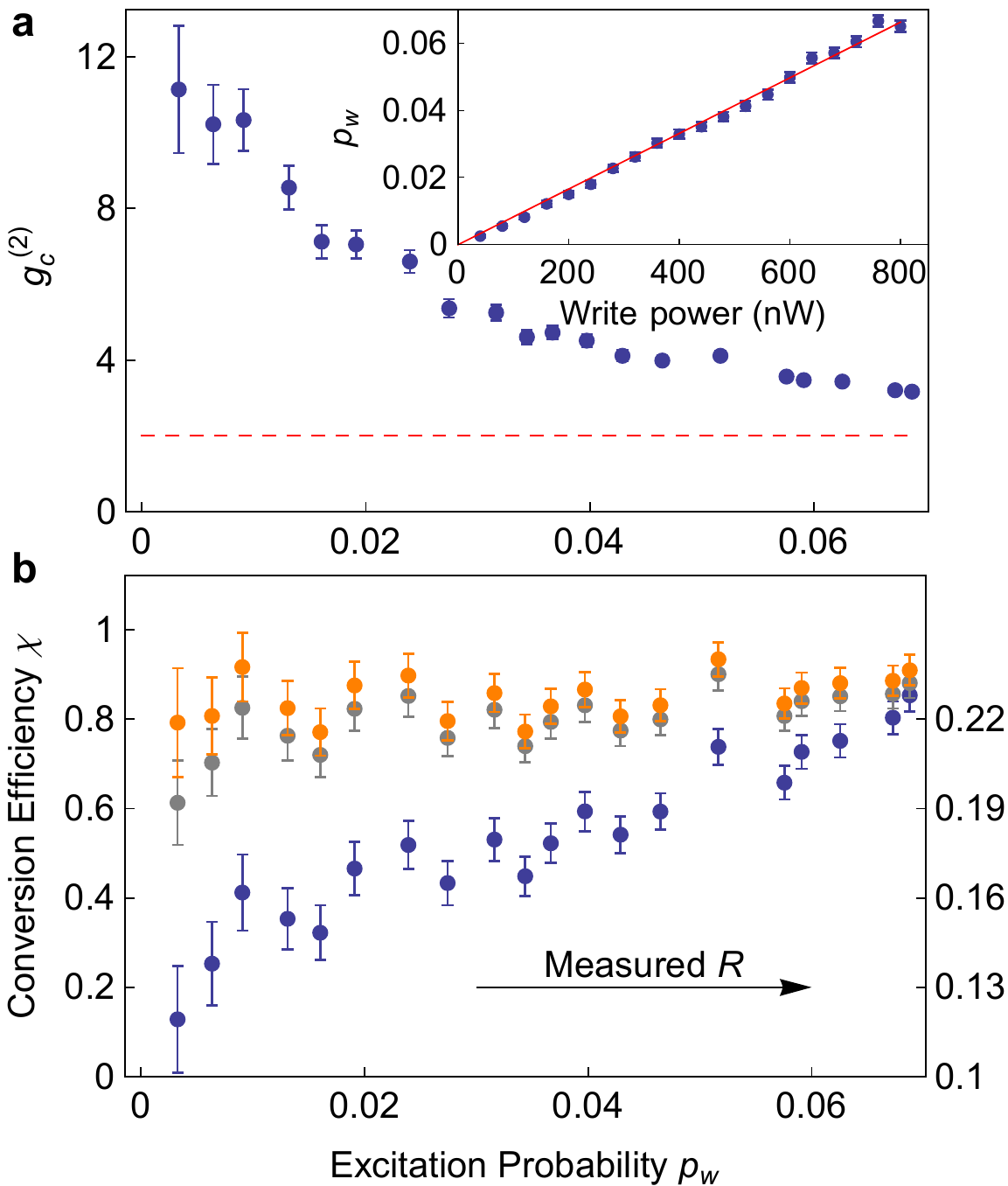}
\caption{\textbf{Influence of the write power.} \textbf{a}, Cross correlation as
a function of the excitation probability, with the inset showing the relation
between excitation probability and the write power. \textbf{b}, Intrinsic conversion efficiency $\protect\chi$ (with $p_{\mathrm{bg}}$ subtraction in orange, without $p_{\mathrm{bg}}$ subtraction in gray) and conditional retrieval efficiency $R$ (blue) as a function of the excitation probability. Data in (\textbf{a}-\textbf{b}) are measured in the correlation mode (see Methods for details) with a read-write delay of 500 ns. Error bars are derived based on the Poisson distribution of the detector counts.}
\label{write}
\end{figure}

We first demonstrate that the use of the ring cavity preserves the quantum
nature of the atomic memory but significantly enhances the retrieval
efficiency. We measure the excitation probability, cross
correlation and physical retrieval efficiency for a series of write pulses
of different intensities, with the results shown in Fig. \ref{write}a-b,
respectively. We find the excitation probability increases linearly with the write
power, indicating that the write process is within the spontaneous regime
all over the measurement range. We measure the cross correlation between the
write-out and read-out photons, which is defined as $g_{\mathrm{c}}^{(2)}=p_{%
\mathrm{w,\,r}}/(p_{\mathrm{w}}\,p_{\mathrm{r}})$ where $p_{\mathrm{w}}$($p_{%
\mathrm{r}}$) denotes the write(read)-out photon probability and $p_{\mathrm{%
w,\,r}}$ the coincidence probability\cite{Kuzmich2003}. As depicted in Fig. %
\ref{write}a, $g_{\mathrm{c}}^{(2)}$ is well above 2, which implies that the
storage is in the quantum regime and the read-out photon is nonclassically
correlated with the write-out photon \cite{Felinto06,Zhao2009}. The probability of detecting a read-out photon conditional on a write-out photon
event is $R=p_{\mathrm{w,\,r}}/p_{\mathrm{w}}$, whose value ranges from 12\%
to 22\% as shown in Fig. \ref{write}b. The calibrated retrieval efficiency
is calculated by taking into account the background noise $p_{\mathrm{bg}}$ (see Methods for details) and the total detection efficiency, as $R^{\mathrm{c}}=R/[\eta _{\mathrm{tot}}(1-p_{\mathrm{bg}}/p_{\mathrm{w}})]$, where the total detection efficiency $\eta _{\mathrm{tot}}=\eta _{\mathrm{esp}}\,\eta _{\mathrm{t}}\,\eta _{\mathrm{spd}}$
consists of the fraction of light that escapes from the output coupler $\eta _{\mathrm{esp}}=71(2)\%$, the propagation efficiency from the cavity to the detectors (D1 and D2 in Fig. \ref{setup}) $\eta _{\mathrm{t}}=39.5(4)\%$, and the quantum efficiency of
single-photon detectors $\eta _{\mathrm{spd}}=62.7\%$. By subtracting the
random coincidence, we obtain the intrinsic retrieval efficiency
$\chi =R^{\mathrm{c}}(1-1/g_{\mathrm{c}}^{(2)})$, which is displayed as a function of the excitation probability in Fig. \ref{write}b. Note that $R^{\mathrm{c}}$ is often
used as the intrinsic retrieval efficiency in previous experiments and here we explicitly distinguish between $R^{\mathrm{c}}$ and  $\chi $. Over the whole measurement range, $\chi $ approximately stays constant. The calculated average value of $\chi $ is 85(3)\%, which is the highest conversion efficiency of a quantum memory working at single photon level so far.

\begin{figure}[hbtp]
\centering
\includegraphics[width=0.8\columnwidth]{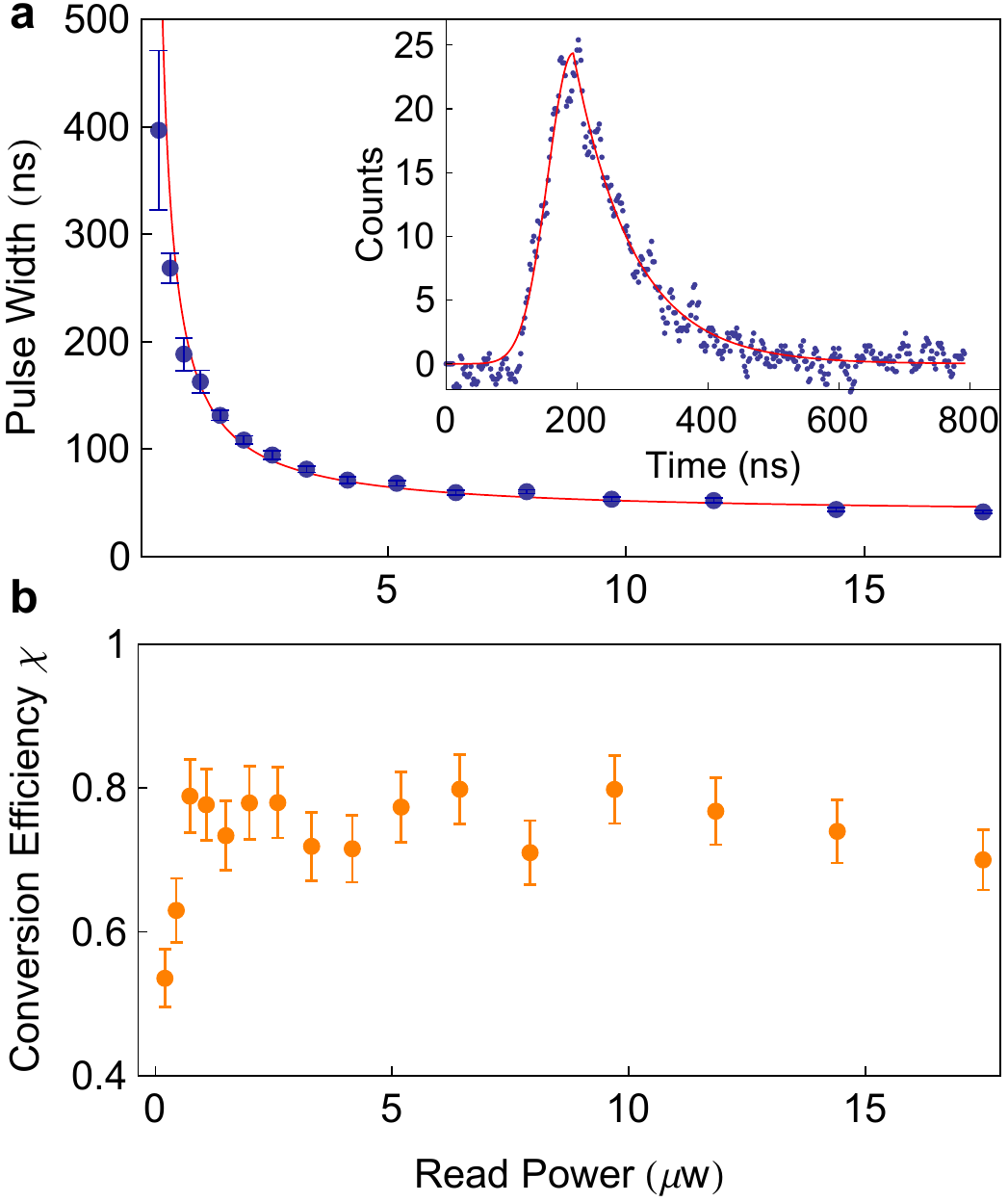}
\caption{\textbf{Influence of the read power.} \textbf{a}, Relation between
the pulse width (FWHM) and the read power, with the inset showing the pulse shape of
the read-out photon with a read intensity of 2 $\protect\mu$W. \textbf{b},
Intrinsic conversion efficiency $\protect\chi$ as a function of the read power.
During the measurements (\textbf{a}-\textbf{b}), the width of the read pulse is set to 700 ns. Error bars are derived based on the Poisson distribution of the
detector counts.}
\label{read}
\end{figure}

Next, we study influence of the read power on the retrieval efficiency. An example of the read-out photon shape is shown in the inset of Fig. \ref{read} a. Experimentally, we find that the pulse width of the read-out photon strongly depends on the read power. With the read power increasing, the pulse width becomes narrower. This is consistent with the EIT theory \cite{Fleischhauer2005} which predicts that the group velocity of the read-out photon should be inversely proportional to the read intensity. The measured relation is plotted in Fig. \ref{read}a and fitted with the function $a/I_{\mathrm{r}}+\tau _{0}$. The fitted result for $\tau _{0}=38.9\pm 1.3$ ns is mainly limited by the cavity decay time (8.9 ns) and the rise time of the read pulse ($\sim $21 ns). We also measure the conversion efficiency $\chi $ as a function of the read power, with the results shown in Fig. \ref {read}b. We note that, except for the first two points, $\chi $ nearly stays constant over the whole measurement range. The variation of the first two points can be explained as a result of the comparable pulse width of the read and read-out pulses which leads to an incomplete read-out process. This implies that we can tune the pulse width for the read-out photon without sacrificing the conversion efficiency. This tunability is of crucial importance for quantum communication and distributed quantum computing \cite{Kimble2008review, Duan2010} since it is usually required to interfere the read-out photon with photons from other physical systems, such as cavity quantum electrodynamics \cite{Kimble2008review}, ion traps \cite{Duan2010} or cavity-enhanced spontaneous parametric down-conversion (SPDC) \cite{Bao2008}, etc.

\begin{figure}[hbtp]
\centering
\includegraphics[width=0.8\columnwidth]{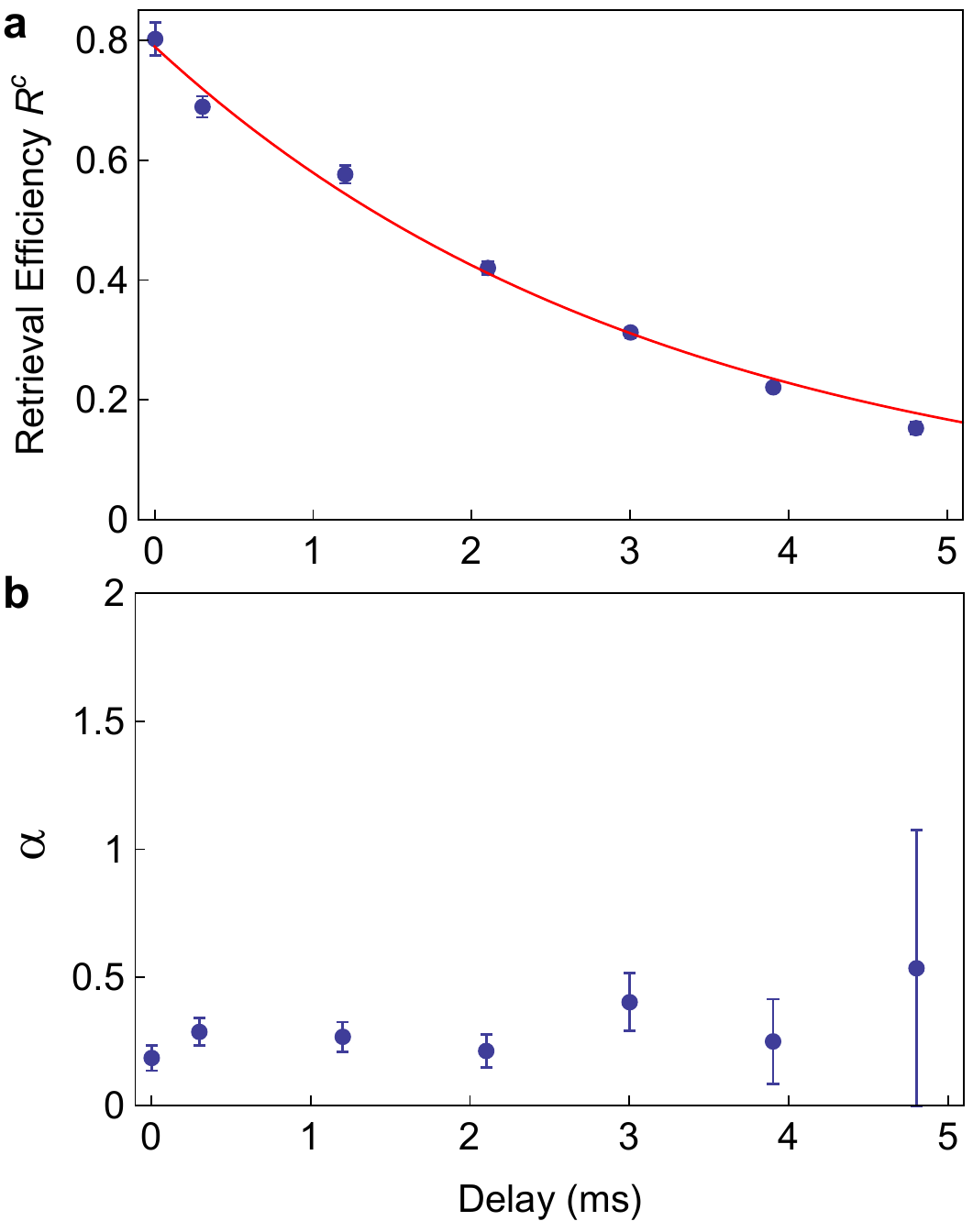}
\caption{\textbf{Long lifetime measurement.} \textbf{a}, Calibrated
retrieval efficiency $R^\mathrm{c}$ as a function of the time delay between read
and write. \textbf{b}, Anti-correlation parameter $\protect\alpha$ of the
conditional read-out photon as a function of the time delay. Error bars are
derived based on the Poisson distribution of the detector counts.}
\label{lifetime}
\end{figure}

Finally we suppress different decoherence mechanisms to achieve an efficient
and long-lived quantum memory. In our experiment, the main sources of
decoherence are inhomogeneous broadening due to residual magnetic fields,
spin wave dephasing, and atoms moving out of the interaction region \cite%
{Zhao2009}. To minimize the decoherence caused by the residual magnetic
field, we employ the clock states $|g\rangle =|F=2,m_{F}=0\rangle $ and $%
|s\rangle =|F=1,m_{F}=0\rangle ,$ which are in the first order immune to the
magnetic field variations. Imperfect pumping to the state $|g\rangle$ gives rise to slight
oscillation of the retrieval efficiency for the short-time ($\sim $10 $\mu $%
s) storage. From the relative oscillation amplitude of $\sim $14\%, we
estimate \cite{ZhaoR2009} a pumping efficiency of $p_{0}\approx 95\%$, by assuming no
occupation in the Zeeman sub-levels $|F=2,m_{F}=\pm 2\rangle $. The spin
wave dephasing is suppressed by selecting only the forward scattered
write-out photons, which results in a long-wavelength spin wave. After
suppressing these two decoherence mechanisms, the remaining dominant
decoherence mechanism is the loss of atoms caused by atoms freely flying out
of the interaction region. We take two measures to minimize this
decoherence. First, we configure the cavity mode in the same direction as the gravity. In this way, the maximal allowed free falling distance
(from the atomic ensemble to the walls of the glass cell) is about 1.5 cm, which
corresponds to a falling time of 55 ms. Second, we increase the beam waist
of the detection mode from a typical value of 100 $\mu $m in previous
experiments \cite{Zhao2009,Yuan2008} to the current value of 200 $\mu $m to
reduce the loss of atoms caused by the free expansion of the ensemble. For our
temperature of about 10 $\mu $K, the estimated lifetime is 4.7 ms.

Experimentally we perform this long-lifetime measurement in a feedback mode (see Methods for details) to save data integration time. In this mode, the write process is repeated until a write-out photon is detected. The read pulse is applied only when a
write-out event has been registered. In Fig. \ref{lifetime}a, we plot the
calibrated retrieval efficiency $R^{\mathrm{c}}$ as a function of the
storage time. Note that the dominant decoherence mechanism for short storage time is caused by the imperfect pumping and residual magnetic field\cite{ZhaoR2009}, and for long storage time is due to atomic motion\cite{Zhao2009}. Here we fit the data to an exponential function $R_{0}^{\mathrm{c}}\,e^{-t/\tau }$ for simplicity, which give the results $R_{0}^{\mathrm{c}}=79(2)\,\%$ and $%
\tau =3.2(1)\,\mathrm{ms}$. We attribute the discrepancy between the
lifetime $\tau $ and previous theoretical estimations to additional heating
during the pumping process. From this fitted value of $R_{0}^{\mathrm{c}}$
and using $p_{\mathrm{r}}=0.0094(2)$ and $R=0.127(3)$ for the first point in
Fig. \ref{lifetime}, we calculate the intrinsic spin wave to photon
conversion efficiency to be $\chi =73(2)\%$. We can also see that the quantum memory maintains a retrieval efficiency in excess of 30\% for about 3 ms. To further prove that the storage is in the quantum regime, we measure the anti-correlation parameter $%
\alpha $ which is defined as $\alpha =p_{12}/(p_{1}\,p_{2})$ \cite%
{Grangier1986}, where $p_{1}$($p_{2}$) refers to the detection probability
of D1(D2) in Fig. \ref{setup}b conditional on a detection event of the
write-out photon, and $p_{12}$ refers to the conditional coincidence
probability between D1 and D2. An ideal single-photon corresponds to $\alpha
=0$, while for classical light $\alpha =1$. In our experiment, $\alpha $ is
measured as a function of the storage lifetime, with the results shown in
Fig. \ref{lifetime}b. All the values are well below the classical bound,
which implies that the quantum nature is conserved.

To conclude, we have realized a high-performance atomic-ensemble
quantum memory inside a ring cavity, featuring the simultaneous achievement
of a high retrieval efficiency and a long storage lifetime. The lifetime of
3.2 ms is 10$^{4}$ times higher than the duration (about 100 ns) of pulsed
optical operations and retrieval efficiency of 73\% is larger than the
threshold (50\%) of loss-tolerant linear optical quantum computation.
The fidelity of the quantum memory is mainly limited by the background noises and can be improved by increasing the extinction ratio of pulse switching and single-photon filtering. This realization also enables us to create single-photons with much higher source efficiency using the technique of conditional quantum evolution \cite{Matsukevich2006}. Further improvement of storage lifetime could be possible by making use of a light-compensated optical lattice \cite{Radnaev2010}. An even further improved retrieval efficiency may be achieved by increasing the pumping efficiency for high optical depths. This demonstration of a high-efficiency and long-lifetime quantum
memory enables the implementation of advanced quantum information tasks
such as connecting different quantum repeater nodes, the purification \cite%
{Pan2001} of remote entanglement, and multi-ensemble entanglement creation.

\subsection*{Methods}

\textbf{Technical details.} The repetition rate of our experiment is $\sim$29 Hz. Within each cycle, the starting 31.5 ms is used for the MOT loading phase, during which we recapture the atoms and recool them. Afterwards, the experimental phase starts, with a duration between 1 and 4.8 ms determined by the storage time. Within the starting 1 ms of each experimental phase, the write cycle repeats with a repetition rate of 154 kHz (in feedback mode). An optical ring cavity is placed outside of the vacuum glass-cell chamber, consisting of one output coupler (OC) with a transmission rate of 8.81(3)\%, two high-reflection (HR) mirrors, and two lenses (22.5 cm). The cavity has a round-trip length of 50.3 cm and a free spectral range (FSR) of 595.1 MHz. The glass cell windows and the lens surfaces are antireflection coated. Within one round trip traveling, there is a $\pi$ phase shift between H and V, which explains why the ratio of ground state splitting over one FSR does not equal to an integer. Beam waist of the cavity mode can be slightly changed by moving the lens positions.

\textbf{Fidelity of the quantum memory.} It can be characterized either by the cross correlation\cite{Zhao2009} $g_{\mathrm{c}}^{(2)}$ or the anti-correlation parameter\cite{ZhaoR2009} $\alpha$. If the atomic ensemble is used as a single-photon source, the anti-correlation parameter $\alpha$ determines the single-photon quality. If the atomic ensemble is used to implement atom-photon entanglement, the violation of Bell's inequality is characterized by $S>2$, where $S$ may be expressed as $S\approx2\sqrt{2}V$ with the visibility $V\approx(g_{\mathrm{c}}^{(2)}-1)/(g_{\mathrm{c}}^{(2)}+1)$. Therefore, $g_{\mathrm{c}}^{(2)}\gg1$ and $\alpha\ll1$ both demonstrate a high-fidelity quantum memory.

\textbf{Cavity enhancement.} The utilization of a ring cavity and configuring it to be quadruple resonant with the write(read) beams and the write-out (read-out) photons offer a series of advantages. First, the intensity of write and read beams inside the cavity is enhanced by a factor of $2F\eta _{\mathrm{esp}}/\pi $, which moderates the
requirement of single-photon filtering a lot. Second, the ring cavity
enhances the emission of the write-out photon into the cavity mode by a
factor of $2F/\pi $. Third and most importantly, the ring cavity enhances the coherent read-out emission by Purcell effect, which gives rise to a maximal retrieval efficiency \cite{Gorshkov2007} of $\chi\approx C/(C+1)$ where the cooperativity $C \propto F \times d$, with $d$ the optical depth of the atomic ensemble. This relation can be understood as the competition between the collective emission of the read-out field into the cavity mode, which is further enhanced by Purcell effect, and spontaneous emission into free space through partial population in the excited state.

\textbf{Phase shift of the atomic ensemble.} During the spin wave creation
and read-out process, most of atoms populate in the state $|g\rangle$.
Therefore, due to close resonance of the write and the read-out photon to
the $|g\rangle$ $\rightarrow$ $|e\rangle$ transition, the atomic ensemble
will give rise to a phase shift to the write beam and
read-out photons. In our experiment this phase shift is measured by
monitoring the leakage signal of the write beam from HR2 as we scan the
locking beam center frequency. In order to compensate this phase shift ($\sim4^\circ$) and achieve the resonance between write(read-out) and the ring cavity mode, we set a frequency offset $\delta$ to the locking beam. Therefore the locking beam is blue detuned by one FSR +
$\delta$ relative to the write in total. To achieve resonance of the read
beam and write-out photons with the ring cavity, we make use of a manual
translation stage which is attached to HR2 to change the FSR, and scan its
position while monitoring the leakage signal of the read beam from HR2.

\textbf{Measurement modes.} We have two different modes for data
measurement, i.e., correlation mode \cite{Zhao2009} and feedback mode \cite{ZhaoR2009}. In the correlation mode, each write pulse is followed by a read pulse with a fixed delay. In
this mode since $p_{\mathrm{w}}$, $p_{\mathrm{r}}$ and $p_{\mathrm{w,r}}$
are measured, we can verify the quantum character of our memory using the
cross correlation $g_{\mathrm{c}}^{(2)}$. In the feedback mode, the write
process repeats until a write-out photon is detected. The read pulse is
only applied conditioned on these events. In this mode, due to the
unavailability of $p_{\mathrm{r}}$, we measure the anti-correlation parameter $\alpha $ to verify the quantum character.

\textbf{Background subtraction.} The background $p_{\mathrm{bg}}$ in the write process
can be separated into two parts. The first part includes the stray light, dark count of single-photon detectors, and the leakage of read and locking beam due to the limited extinction ratio of the AOMs, etc, which are not dependent on the write power and give an overall constant background of 0.0006(1). The second part is mainly due to the leakage of the write laser, which increases linearly as a function of the write power. The write leakage induced background contributes to $\sim2.3\%$ for $p_{\mathrm{w}}$. Since the write leakage can be further decreased by using more stages of frequency filtering, it is reasonable to subtract this background, which will also enable a more accurate estimate of $\chi$.

\bibliography{cavref}

\renewcommand\subsection[1]{
\vspace{\baselineskip}
\textbf{#1}
\vspace{0.5\baselineskip}
}

\subsection{Acknowledgement}

This work was supported by the European Commission through the ERC Grant,
the STREP project HIP, the CAS, the NNSFC, and the National Fundamental
Research Program (Grant No. 2011CB921300) of China.

\subsection{Author Contributions}

X.-H.B., A.D., B.Z. and J.-W.P. conceived and designed the experiment. A.D., P.D., A.R., T.S. and X.-H.B. built the setup. X.-H.B., A.R., P.D. and J.R. carried out the experiment. X.-H.B., A.R., L.L., N.-L.L. and B.Z. analyzed the data. X.-H.B. wrote the paper with substantial contributions by B.Z. and all authors. J.-W.P. supervised the project.

\subsection{Competing Interests}

The authors declare that they have no competing financial interests.

\subsection{Correspondence}

Correspondence and requests for materials should be addressed to J.-W.P.
(e-mail: pan@ustc.edu.cn) or B.Z. (e-mail: bozhao@ustc.edu.cn).

\end{document}